\documentclass[submitting]{nst}

\usepackage{graphicx}     

\usepackage{xcolor}
\usepackage[normalem]{ulem}

\usepackage{booktabs}     
\usepackage{multirow}      
\usepackage{array}        
\usepackage{dcolumn}

\usepackage{amsmath,amssymb,amsfonts}
\usepackage{siunitx}
\usepackage{upgreek}

\usepackage{subfigure}

\begin{document}

\title{An improved linear Boltzmann transport model for hadron and jet suppression in ultrarelativistic heavy-ion collisions}

\author{Yichao Dang}
\affiliation{Institute of Frontier and Interdisciplinary Science, Shandong University, Qingdao, Shandong 266237, China}

\author{Wen-Jing Xing}
\email{wenjing.xing@usc.edu.cn}
\affiliation{School of Nuclear Science and Technology, University of South China, Hengyang, Hunan 421001, China}

\author{Shanshan Cao}
\email{shanshan.cao@sdu.edu.cn}
\affiliation{Institute of Frontier and Interdisciplinary Science, Shandong University, Qingdao, Shandong 266237, China}

\author{Guang-You Qin}
\email{guangyou.qin@ccnu.edu.cn}
\affiliation{Institute of Particle Physics and Key Laboratory of Quark and Lepton Physics (MOE), Central China Normal University, Wuhan, 430079, China}

\date{\today}

%%%%%%%%%%%%%%%%%%%%%%%%%%%%%%%%%%%%%%%%%%%%%%%%%%%%%%%%%%%%%%%%%%%%%

\begin{abstract}
Jets serve as powerful tomographic probes of the quark-gluon plasma (QGP) created in relativistic heavy-ion collisions. While the expanding landscape of jet observables reveals multi-faceted aspects of jet-medium interactions, a precise and simultaneous description of the nuclear modification factors of hadrons and full jets remains a challenge for theoretical models. In this work, we present two essential improvements to the linear Boltzmann transport (LBT) model to bridge this gap. First, instead of implementing in-medium parton transport after vacuum parton showers complete, we introduce a medium scale at which in-medium parton transport is inserted into the vacuum parton showers, providing a more physical picture of parton-QGP interactions. Second, we incorporate color flow information into the LBT model, enabling string connections between partons whose configurations are correlated with the medium-modified parton showers before hadronization. We demonstrate that both improvements alter the predicted ratio of hadron to jet quenching, leading to a satisfactory unified description of the nuclear modification factors of hadrons and jets with different flavors.
\end{abstract}
\keywords{relativistic heavy-ion collisions, quark-gluon plasma, jet quenching}
\maketitle

%%%%%%%%%%%%%%%%%%%%%%%%%%%%%%%%%%%%%%%%%%%%%%%%%%%%%%%%%%%%%%%%%%%%%

\section{Introduction}
\label{sec:introduction}

High-energy nuclear collisions conducted at the Relativistic Heavy-Ion Collider (RHIC) and the Large Hadron Collider (LHC) provide a unique opportunity to study the color-deconfined Quantum Chromodynamics (QCD) matter, known as the quark-gluon plasma (QGP)~\cite{Busza:2018rrf,Elfner:2022iae,Harris:2024aov,Chen:2026gka}. A smoking gun signature of the QGP formation is the significant suppression of the high-transverse-momentum ($p_\mathrm{T}$) hadron yields in nucleus-nucleus (A$+$A) collisions relative to those in proton-proton ($p+p$) collisions~\cite{PHENIX:2012jha,CMS:2012aa,ALICE:2012aqc}, indicating strong energy loss of high-$p_\mathrm{T}$ quarks and gluons inside the QGP before fragmenting into hadrons. This phenomenon is called jet quenching and serves as a powerful probe of the microscopic structures of the QGP~\cite{Wang:1992qdg,Qin:2015srf,Majumder:2010qh,Cao:2020wlm,Wang:2025lct,Mehtar-Tani:2025rty}. The jet quenching parameter $\hat{q}$, characterizing the amount of in-medium deflection experienced by high-$p_\mathrm{T}$ particles, is found over an order of magnitude larger in the QGP than in cold nuclei~\cite{JET:2013cls,JETSCAPE:2021ehl}, suggesting the dense partonic degrees of freedom inside the QGP. 

Various parton energy loss formalisms have been developed based on different assumptions of the QGP properties and the kinematics of jet partons and their emitted gluons~\cite{Gyulassy:1993hr,Wang:1994fx,Gyulassy:2000fs,Gyulassy:2000er,Djordjevic:2004nq,Baier:1996kr,Baier:1996sk,Zakharov:1996fv,Zakharov:1997uu,Zakharov:1998sv,Wiedemann:2000za,Arnold:2002ja,Guo:2000nz,Wang:2001ifa,Majumder:2009ge,Sirimanna:2021sqx,Chesler:2014jva,Chen:2024epd}, and sophisticated Monte-Carlo event generators have been built to realize these energy loss formalisms in the dynamical environment of heavy-ion collisions~\cite{Cao:2024pxc}. There are two main categories of event generators for jet quenching. The first category introduces medium modifications into the vacuum showers of high-virtuality partons; examples include the {QPythia}~\cite{Armesto:2009fj}, {JEWEL}~\cite{Zapp:2008gi,Zapp:2012ak}, {Hybrid}~\cite{Casalderrey-Solana:2014bpa} and {MATTER}~\cite{Cao:2017qpx} models. The second category comprises transport models, which start from the final states of vacuum parton showers and simulate subsequent parton interactions with the QGP at low virtuality scales, such as the LBT~\cite{Luo:2023nsi}, {MARTINI}~\cite{Schenke:2009gb} and {LIDO}~\cite{Ke:2018jem,Ke:2020clc} models. The {JETSCAPE} framework is developed to combine different energy loss models into a unified approach~\cite{JETSCAPE:2017eso,Putschke:2019yrg}. With the advances in both theoretical calculations and experimental measurements over the past two decades, studies on jet quenching have been extended from single inclusive hadron suppression~\cite{Arnold:2001ba,Wang:2001ifa,Salgado:2003gb,Vitev:2002pf,Vitev:2004bh,Dainese:2004te,Armesto:2005iq,Wicks:2005gt,Bass:2008rv,Armesto:2009zi,Marquet:2009eq,Chen:2010te,Majumder:2010ik,Renk:2010mf,Renk:2011gj,Horowitz:2011gd,Chen:2011vt,Cao:2017hhk} to medium modification of di-hadron~\cite{Majumder:2004pt,Zhang:2007ja,Renk:2008xq,Cao:2015cba} and $\gamma$-hadron~\cite{Zhang:2009rn,Qin:2009bk,Wang:2013cia,Chen:2017zte} correlations. Studies of hadronic observables have also been extended to full-jet~\cite{Qin:2010mn,Dai:2012am,Wang:2013cia,Chang:2016gjp,Casalderrey-Solana:2014bpa,JETSCAPE:2022jer}, which are clusters of particles at various energy scales, and thus are sensitive to not only the energy loss of high-energy (hard) partons, but also the energy flow carried by low-energy (soft) partons. As a result, apart from medium modification of hard partons, tremendous efforts have also been devoted to understanding how hard partons modify the QGP evolution, leading to energy redistribution within a given jet cone. The latter is known as jet-induced medium excitation, or medium response~\cite{Cao:2022odi,Yang:2025lii}, which constitutes a crucial part of jet-medium interactions and affects almost all observables of full jets, such as the yield suppression and collective flow of jets~\cite{He:2018xjv,He:2022evt}, the transverse energy distribution relative to jet axes (jet shape)~\cite{Tachibana:2017syd,Casalderrey-Solana:2016jvj,KunnawalkamElayavalli:2017hxo,Luo:2018pto,Ke:2020clc}, the longitudinal momentum distributions of jet constituents (jet fragmentation function or splitting function)~\cite{KunnawalkamElayavalli:2017hxo,Chen:2017zte,Chen:2020tbl,Ke:2020clc,Milhano:2017nzm,Duan:2025wsy}, two- or multi-point energy correlators of jets~\cite{Yang:2023dwc,Xing:2024yrb,Barata:2024ukm,Bossi:2024qho}, energy depletion in the direction opposite to jet propagation~\cite{Chen:2017zte,Yang:2021qtl,Yang:2022nei,Yang:2025dqu}, and hadron chemistry inside medium-modified jets~\cite{Chen:2021rrp,Luo:2021voy,Sirimanna:2022zje,Luo:2024xog}. 

While the abundant jet observables above offer multi-faceted insights into jet-medium interactions, from which properties of both jets and the QGP have been extracted~\cite{Feal:2019xfl,Xie:2022ght,Xie:2024xbn,Karmakar:2023ity,Liu:2023rfi}, it remains necessary to re-examine whether the simplest quantity of jet quenching, the nuclear modification factor ($R_\mathrm{AA}$), has been precisely understood. The $R_\mathrm{AA}$ factor is defined as the ratio of particle or jet spectrum in A$+$A collisions to that in $p+p$ collisions, properly normalized by the number of binary nucleon-nucleon collisions. While many theoretical models provide satisfactory descriptions of the hadron and jet $R_\mathrm{AA}$'s individually, only a few achieve a simultaneous description of both. By applying a global fit or the Bayesian inference, simultaneous descriptions of the hadron and jet $R_\mathrm{AA}$'s have been obtained within the {Hybrid} model~\cite{Casalderrey-Solana:2018wrw} and the {Lido} model~\cite{Ke:2020clc}. On the other hand, even with the Bayesian calibration, the recent {JETSCAPE} study still encounters tension between the hadron and jet $R_\mathrm{AA}$'s~\cite{JETSCAPE:2024cqe}. Similarly, within the linear Boltzmann transport (LBT) model we developed earlier, different values of the strong coupling coefficient $\alpha_\mathrm{s}$ were used for studies on nuclear modification of hadrons~\cite{Xing:2019xae} and jets~\cite{He:2018xjv,He:2022evt}. Therefore, while state-of-the-art statistical tools are powerful for parameter fitting, they are insufficient. A precise understanding requires dissecting the key physics ingredients that affect the quenching of hadrons relative to jets. This is the main purpose of the present work. Note that a unified framework of the hadron and jet quenching also provides a baseline for understanding the recent puzzling acoplanarity of hadron-triggered jets observed at RHIC~\cite{STAR:2025yhg} and the LHC~\cite{ALICE:2023qve}.

In this work, we present two key improvements to the LBT model. First, instead of initiating parton transport after the completion of the vacuum parton showers, we interrupt the {Pythia} vacuum shower~\cite{Sjostrand:2006za,Sjostrand:2014zea} at a scale characteristic of the QGP and insert the parton transport in between. This creates a more realistic picture: highly virtual jet partons first evolve toward the medium scale via vacuum-like splittings, then scatter with the QGP with their virtuality held at the medium scale, and finally exit the medium and undergo vacuum showers again towards the hadronization scale before turning into hadrons. We demonstrate that introducing this medium scale significantly alters the predicted ratio of hadron to jet quenching. Second, we incorporate color flow information into the LBT model for both elastic and inelastic scatterings. This extension enables the use of the {Pythia} string fragmentation for hadronization of medium-modified partons, allowing us, for the first time, to study full jets at the hadronic level within the LBT framework. We find that the final hadron-to-jet spectrum depends on this color information. The primary goal of this work is to identify essential physics ingredients for a unified description of the hadron and jet $R_\mathrm{AA}$'s, rather than to perform a precise parameter tuning to data. The rest of this paper is organized as follows. In Sec.~\ref{sec:model}, we detail the improved Monte-Carlo framework for jet production, in-medium evolution, and hadronization. In Sec.~\ref{sec:mediumscale}, we analyze the impact of the medium scale and color flow on the quenching of hadrons and jets. A summary is given in Sec.~\ref{sec:summary}.

\section{Jet parton production, evolution, and hadronization}
\label{sec:model}

We use the {Pythia~8} event generator~\cite{Sjostrand:2014zea} to simulate the hard parton production from nucleon-nucleon collisions and the subsequent vacuum parton showers. For nucleus-nucleus collisions, the positions of the binary nucleon-nucleon scattering vertices are sampled according to the Monte-Carlo Glauber model \cite{Miller:2007ri}. Unlike in our earlier studies~\cite{Zhang:2022ctd,Luo:2023nsi} where jet partons start interacting with the QGP medium after they reach the hadronization scale in Pythia ($Q_\mathrm{h}=0.5$~GeV), we introduce a medium scale ($Q_\mathrm{M}$) at which the interactions commence. The formation time of a parton ($\tau_\mathrm{f}$) is defined as the sum of the splitting times of its ancestors before $Q_\mathrm{M}$ is reached~\cite{Zhang:2022ctd}. For an $i \rightarrow jk$ splitting process, where the virtualities of $j$ and $k$ are much smaller than that of $i$, the splitting time is given by~\cite{Adil:2006ra}:
\begin{equation}
\label{eq:tauijk}
\tau_{i \rightarrow jk} = \frac{2z(1-z)E_i}{k_\perp^2+(1-z)m_j^2+zm_k^2-z(1-z)m_i^2},
\end{equation}
with $E_i$ the energy of the parent parton $i$, $k_\perp$ the transverse momentum of daughter partons $j$ and $k$ relative to $i$, $z$ the fractional energy of $i$ taken by $j$, and $m_{i,j,k}$ the rest masses of $i$, $j$, and $k$.
Before its formation time $\tau_\mathrm{f}$, a parton is assumed to stream freely from the production vertex of its earliest ancestor. The jet parton's interaction with the QGP should also start after the initial time of the hydrodynamic evolution of the QGP ($\tau_0 = 0.6$~fm). As discussed in Refs.~\cite{Zhang:2022ctd,JETSCAPE:2017eso}, $\tau_\mathrm{f}$ can exceed $\tau_0$ and is sensitive to both the parton energy and the medium scale: soft partons usually reach a given scale later than hard partons, and raising $Q_\mathrm{M}$ shortens the formation time. Therefore, varying $Q_\mathrm{M}$ affects the relative magnitude of energy loss between soft and hard partons inside the QGP. 

After both $\tau_0$ and $\tau_\mathrm{f}$ are reached, we use the LBT~\cite{Cao:2016gvr,Luo:2023nsi} model to simulate the scatterings of jet partons with the QGP medium. In the local rest frame of the medium, the phase space distribution of jet partons, $f_a(\vec{x}_a,\vec{p}_a,t)$, evolves according to the Boltzmann equation as
\begin{equation}
    p_a\cdot\partial f_a = E_a\left[C^\mathrm{el}(f_a)+C^\mathrm{inel}(f_a)\right],
\end{equation}
with $p_a=(E_a, \vec{p}_a)$ the four-momentum of the jet parton $a$. On the right hand side, $C^{\text{el}}(f_a)$ and $C^{\text{inel}}(f_a)$ represent the collision integrals of elastic and inelastic scattering processes, respectively. From the former, one may extract the energy- and temperature-dependent elastic scattering rate of a jet parton as 
\begin{equation}
\label{eq:elRate}
    \begin{aligned}
    \Gamma_a^\mathrm{el}(E_a&,T) =\sum_{b,(cd)}\frac{\gamma_b}{2E_a}\int\prod_{i=b,c,d}\frac{d^3p_i}{E_i(2\pi)^3}f_b(E_b,T)\\
    &\times[1\pm f_c(E_c,T)][1\pm f_d(E_d,T)]S_2(\hat{s},\hat{t},\hat{u})\\
    &\times(2\pi)^4\delta^{(4)}(p_a+p_b-p_c-p_d)|\mathcal{M}_{ab\to cd}|^2,
    \end{aligned}  
\end{equation}
in which the sum is over all possible scattering channels $ab \rightarrow cd$, with $b$ representing a medium constituent parton, $c$ and $d$ the final state partons of $a$ and $b$, respectively. In addition, $\gamma_b$ denotes the spin-color degeneracy of parton $b$, and thermal (Bose or Fermi) distributions are taken for $f_b$, $f_c$, and $f_d$. In this work, we assume zero masses for light flavor quarks and gluons, and take $m_c = \SI{1.3}{GeV}$ for charm quarks and $m_b = \SI{4.2}{GeV}$ for bottom quarks. Leading order matrix elements $|\mathcal{M}_{ab\to cd} (\hat{s},\hat{t},\hat{u})|^2$~\cite{Auvinen:2009qm} are adopted, with their possible divergence at $\hat{s}, \hat{t}, \hat{u} \rightarrow 0$ regulated by a double-$\theta$ function $S_2(\hat{s},\hat{t},\hat{u}) = \theta(\hat{s}\geq2\mu_{\mathrm{D}}^{2})\theta(-\hat{s}+\mu_{\mathrm{D}}^{2}\leq\hat{t}\leq-\mu_{\mathrm{D}}^{2})$, where $\hat{s},\hat{t},\hat{u}$ are the Mandelstam variables, $\mu_\mathrm{D}^2=4\pi\alpha_\mathrm{s}T^2(N_c+N_f/2)/3$ is the Debye screening mass, with $\alpha_\mathrm{s}$ the strong coupling parameter, $N_c$ and $N_f$ the numbers of colors and flavors, respectively. 

Scatterings with the medium can increase the virtuality of jet partons and induce their additional splittings. These are known as medium-induced gluon emissions, or inelastic scatterings in the LBT model. Hence, the inelastic scattering rate can be related to the average number of gluon emissions per unit time ($t$) as 
\begin{equation}
    \label{eq:inelRate}
    \Gamma_a^{\mathrm{inel}}(E_a,T,t)  =\int dzdk_\perp^2\frac{1}{1+\delta^{ag}}\frac{dN_g^a}{dzdk_\perp^2dt},
\end{equation}
in which $z$ and $k_\perp$ are the fractional energy and transverse momentum of an emitted gluon relative to its parent parton. The gluon spectrum is taken from the higher-twist energy loss calculation~\cite{Wang:2001ifa,Zhang:2003wk,Majumder:2009ge} as
\begin{equation}
    \frac{dN_g^a}{dzdk_\perp^2dt}=\frac{2\alpha_\mathrm{s}C_A\hat{q}_a P_a(z)k_\perp^4}{\pi (k_\perp^2+z^2 m_a^2)^4} \sin^2\left(\frac{t-t_\mathrm{init}}{2\tau^\prime_\mathrm{f}}\right),
\end{equation}
where $C_A=N_c=3$, $m_a$ is the jet parton mass, $P_a(z)$ is the splitting function given by 
\begin{align}
&P_{q\rightarrow qg}(z)=\frac{(1-z)[1+(1-z)^2]}{z},\\
&P_{g\rightarrow gg}(z)=\frac{2(1-z+z^2)^3}{z(1-z)}.
\end{align}
Note that the Kronecker delta function in Eq.~(\ref{eq:inelRate}) generates a $1/2$ factor when converting the final state gluon numbers of a $g \rightarrow gg$ process into the scattering rate of the parent gluon. Furthermore, $t_\mathrm{init}$ denotes the production time of the jet parton, or the time when its previous splitting happens; and according to Eq.~(\ref{eq:tauijk}), the formation time of a massless medium-induced gluon reads 
\begin{equation}
\tau^\prime_\mathrm{f}=\frac{2E_a z(1-z)}{k_\perp^2+z^2 m_a^2}. 
\end{equation}
The jet transport coefficient $\hat{q}_a$ characterizes the transverse momentum broadening square of the jet parton $a$ per unit length (time), $\hat{q}_a = d \langle p_{\perp a}^2\rangle/dt $, which can be evaluated from Eq.~(\ref{eq:elRate}) by including a factor of $p_{\perp a}^2=[\vec{p}_c-(\vec{p}_c \cdot \hat{p}_a)\hat{p}_a]^2$ in its integrand. In principle, the strong coupling coefficient $\alpha_\mathrm{s}$ is the only parameter of the LBT model.

Following our earlier studies~\cite{Cao:2017hhk,Xing:2019xae}, two forms of $\alpha_\mathrm{s}$ are implemented in this work. The coupling coefficient at vertices connected to a high-energy jet parton is assumed to run with both the parton energy and the medium temperature as $\alpha_s = 4\pi / [9 \ln(2ET / \Lambda^2)]$ with $\Lambda = \SI{0.2}{GeV}$. For vertices attached to thermal partons, a fixed $\alpha_\mathrm{s}$ is used. The value of this fixed $\alpha_\mathrm{s}$ is treated as a model parameter to be constrained by the jet quenching data. While the perturbative calculation has been shown successful in describing the nuclear modification of high-$p_\mathrm{T}$ hadrons~\cite{Xing:2019xae}, it is necessary to introduce non-perturbative corrections to scatterings between low energy jet partons and the QGP, especially for describing the nuclear modification of low-$p_\mathrm{T}$ heavy flavor hadrons. In the present work, we introduce a momentum-dependent $K_p$ factor, $K_p = 1+ A_p \exp (-|\vec{p}|^2 / 2\sigma_p^2)$, to enhance the jet transport coefficient $\hat{q}$ at low momentum. The amplitude and width parameters are set as $A_p = 5$ and $\sigma_p = \SI{5}{GeV}$ based on an earlier fit to the $D$ meson data at RHIC and the LHC~\cite{Cao:2016gvr}. Although the non-perturbative corrections for light partons, charm quarks, and bottom quarks are expected to differ, we adopt the same $K_p$ factor for all of them as a minimal and simplifying assumption. More rigorous treatments of non-perturbative effects, such as those developed for heavy quarks~\cite{Xing:2021xwc,Dang:2023tmb,He:2011qa}, will be incorporated in our future efforts.

Based on the rates of elastic and inelastic scatterings, we apply the Monte-Carlo method to simulate the jet parton evolution through the QGP. The local temperature and flow velocity profiles of the QGP medium is generated by the (3+1)-dimensional {CLVisc} hydrodynamic model \cite{Pang:2018zzo,Wu:2021fjf}. At each time step, jet partons are first boosted into the local rest frame of the medium, in which their scatterings are simulated. Afterwards, they are boosted back to the global frame and propagate to the next time step. Details about numerical implementations can be found in Ref.~\cite{Luo:2023nsi}. In LBT, we track not only the jet partons fed from {Pythia} and their emitted gluons, but also the thermal partons that are scattered out of the medium background by jet partons, named ``recoil partons", and the energy holes left inside the medium, named ``negative partons" or ``back-reaction". Recoil and negative partons constitute jet-induced medium excitation. Recoil partons and medium-induced gluons are allowed to re-scatter with the medium in the same way as the jet partons do. For convenience, we call jet partons, medium-induced gluons, and recoil partons ``positive partons" in our following discussions. We assume the virtualities of jet partons maintain $Q_\mathrm{M}$ inside the QGP due to the balance between virtuality gain from scatterings and virtuality loss from medium-induced splittings. Medium-induced gluons, recoil partons and negative partons are set on mass shells once they are produced. After these partons travel outside QGP medium, i.e., to locations with temperature below $T_\mathrm{pc}=165$~MeV, they are transferred back to {Pythia~8}. Jet partons further evolve from $Q_\mathrm{M}$ to $Q_\mathrm{h}$ via vacuum showers, after which they hadronize together with other positive partons via string fragmentation into ``positive hadrons". Negative partons are connected into strings by themselves and are converted into ``negative hadrons", whose contributions need to be subtracted from all jet observables calculated using positive hadrons.

\begin{figure}[tbp!]
    \centering
    \includegraphics[width=0.85\linewidth]{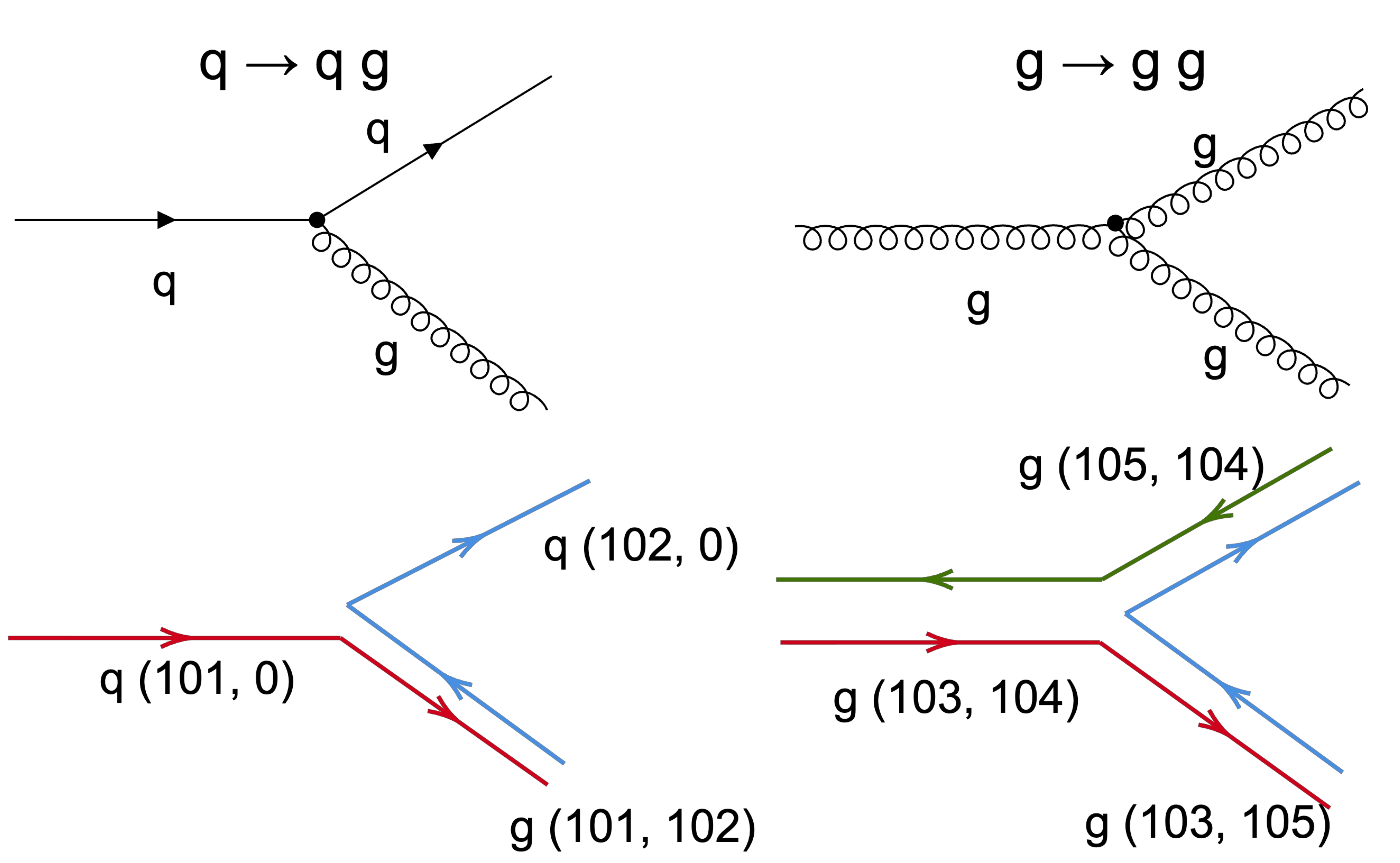}
    \caption{(Color online) Schematic illustration of color flows in $1 \rightarrow 2$ parton splitting processes. The numbers indicate color indices, and the arrows pointing to the left represent anti-colors.}
    \label{fig:split}
  \end{figure}

  \begin{figure}[tbp!]
    \centering
    \includegraphics[width=0.85\linewidth]{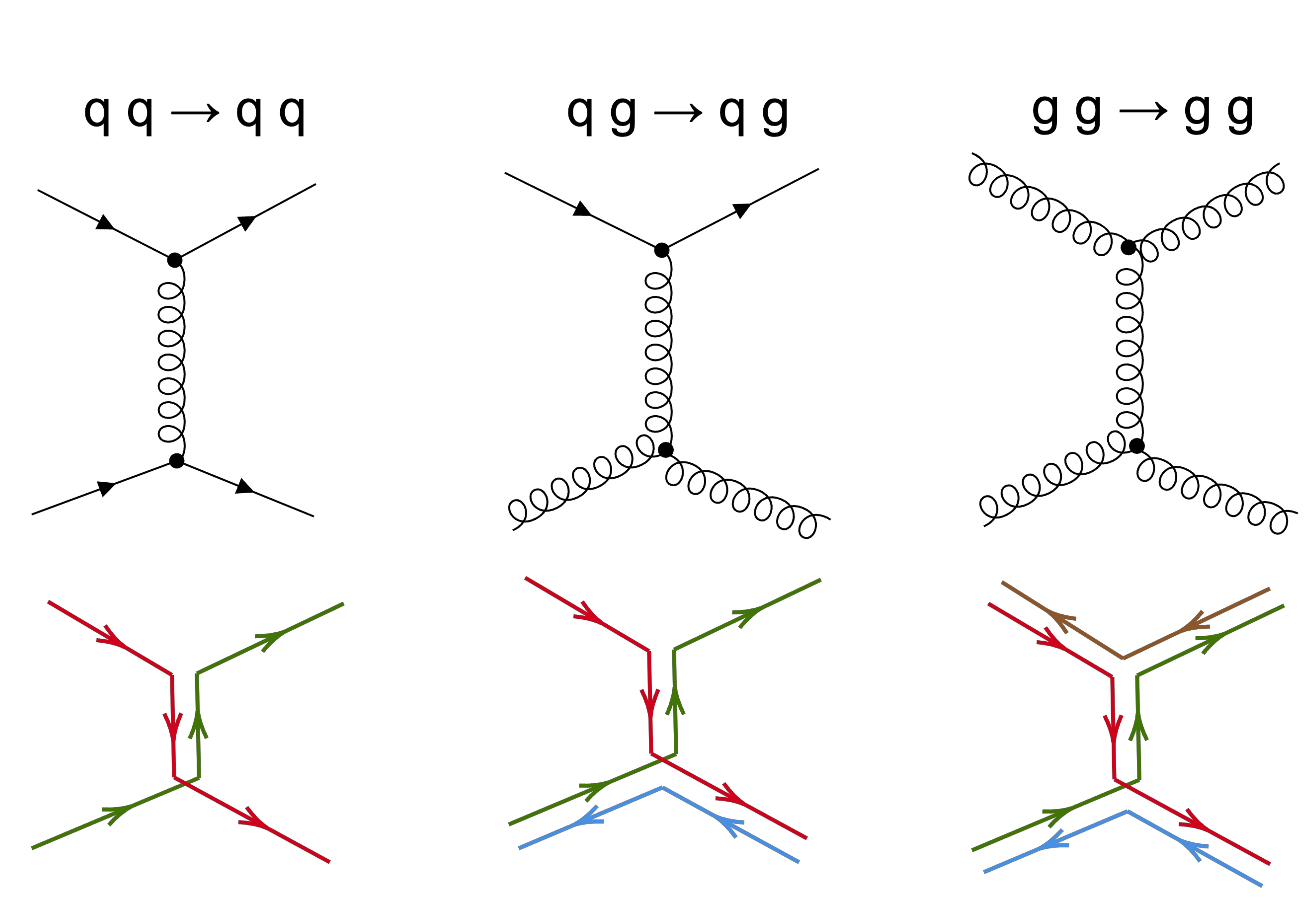}
    \includegraphics[width=0.85\linewidth]{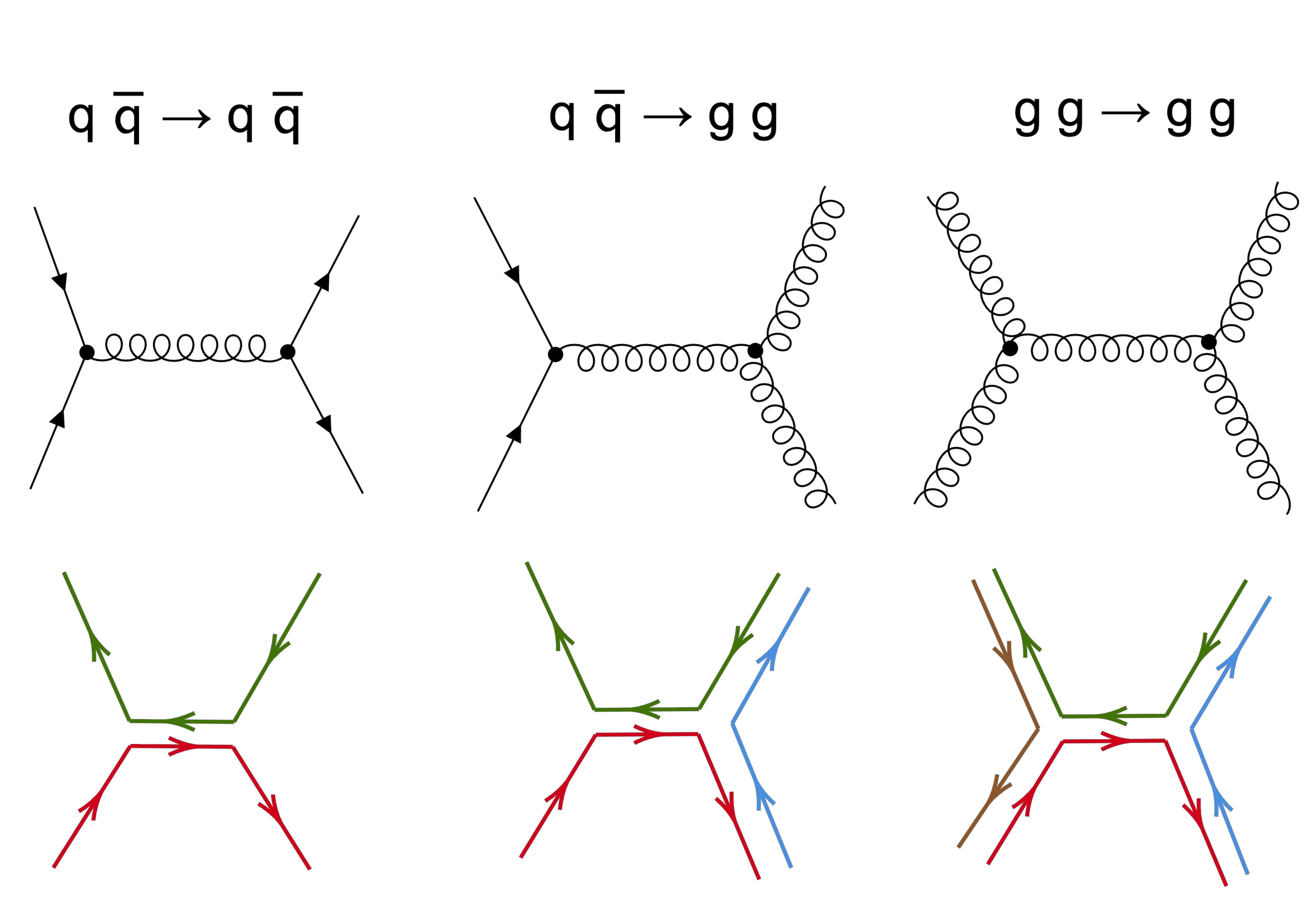}
    \caption{(Color online) Schematic illustration of color flows in $2 \rightarrow 2$ parton scattering processes.}
    \label{fig:scatter}
  \end{figure}

In order to implement vacuum showers and string fragmentation on medium-modified partons in {Pythia~8}, the color information of these partons needs to be properly tracked while they scatter with the medium. We apply the large-$N_c$ color scheme~\cite{tHooft:1973alw} as used in {Pythia~8}, where quarks (anti-quarks) carry color (anti-color) charges, and gluons carry both. The color (anti-color) charges are represented by numerical labels, and color-anti-color pairs with matching labels are connected into strings. Since in LBT, an inelastic scattering process is factorized into an elastic ($2 \rightarrow 2$) scattering followed by one or several ($1 \rightarrow 2$) parton splittings, we focus on the color flows in these two processes below.

Medium-induced gluon emissions correspond to $q \rightarrow qg$ and $g \rightarrow gg$ splittings, as illustrated in Fig.~\ref{fig:split}. We follow the color scheme of vacuum parton showers in {Pythia~8} to track the color flows in these splittings. In Fig.~\ref{fig:split}, each colored line is accompanied by a numerical index, with the flow direction indicating color (rightward) or anti-color (leftward). For the $q \to q  g$ process, the color of the initial quark (101 as an example) is inherited by the emitted gluon. Meanwhile, a new color (102) is generated for the final-state quark, with its associated anti-color assigned to the emitted gluon. In this way, the quark changes its color state through the emission of a gluon. For the $g \to gg$ process, the color and anti-color of the initial gluon are inherited by the two final-state gluons respectively, and an additional color-anti-color line is generated to connect the two final-state gluons. 

\begin{figure}[tbp!]
    \centering
    \includegraphics[width=0.95\linewidth]{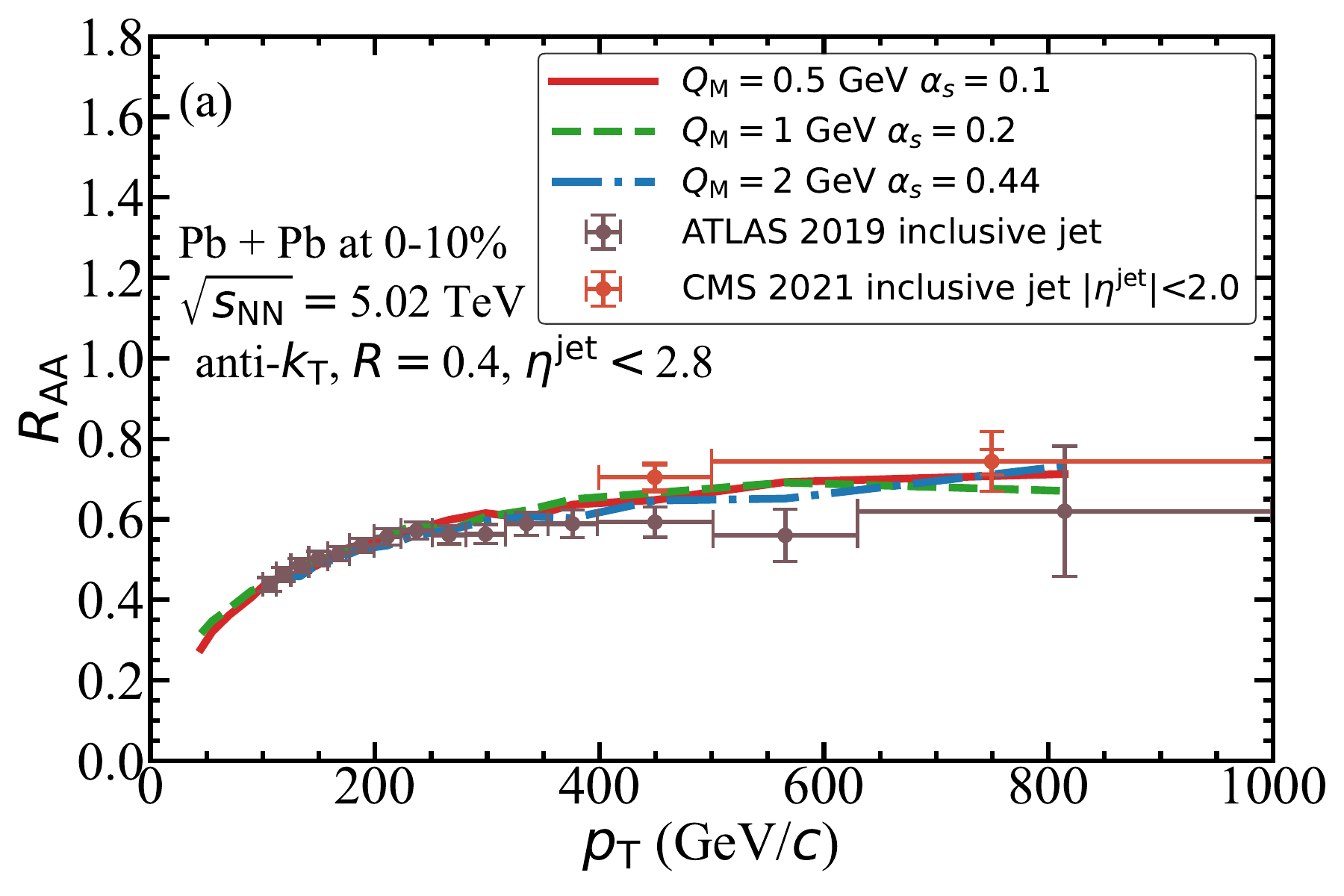}
    \includegraphics[width=0.95\linewidth]{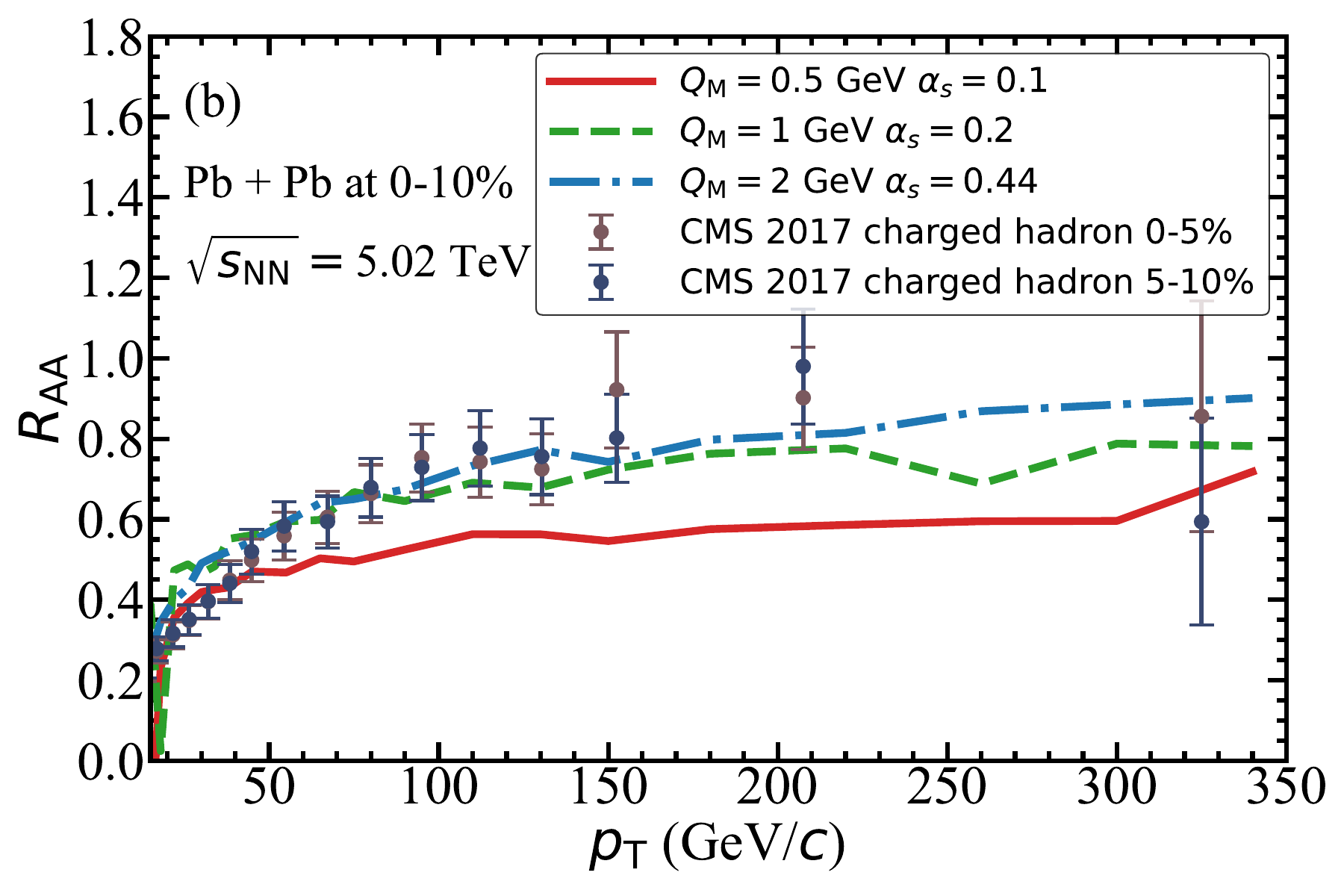}
    \caption{(Color online) The nuclear modification factors of (a) inclusive jets and (b) charged hadrons in central Pb$+$Pb collisions at $\sqrt{s_{\mathrm{NN}}}=\SI{5.02}{TeV}$, compared between using different medium scales. Experimental data are taken from the CMS~\cite{CMS:2021vui,CMS:2016xef} and ATLAS~\cite{Aaboud:2018twu} Collaborations.}
    \label{fig:QMeffect}
\end{figure}

\begin{figure*}[t!]
    \centering
    
    % 第一行：Hadrons
    \subfigure{
        \label{fig:charged_hadron}
        \includegraphics[width=0.33\textwidth]{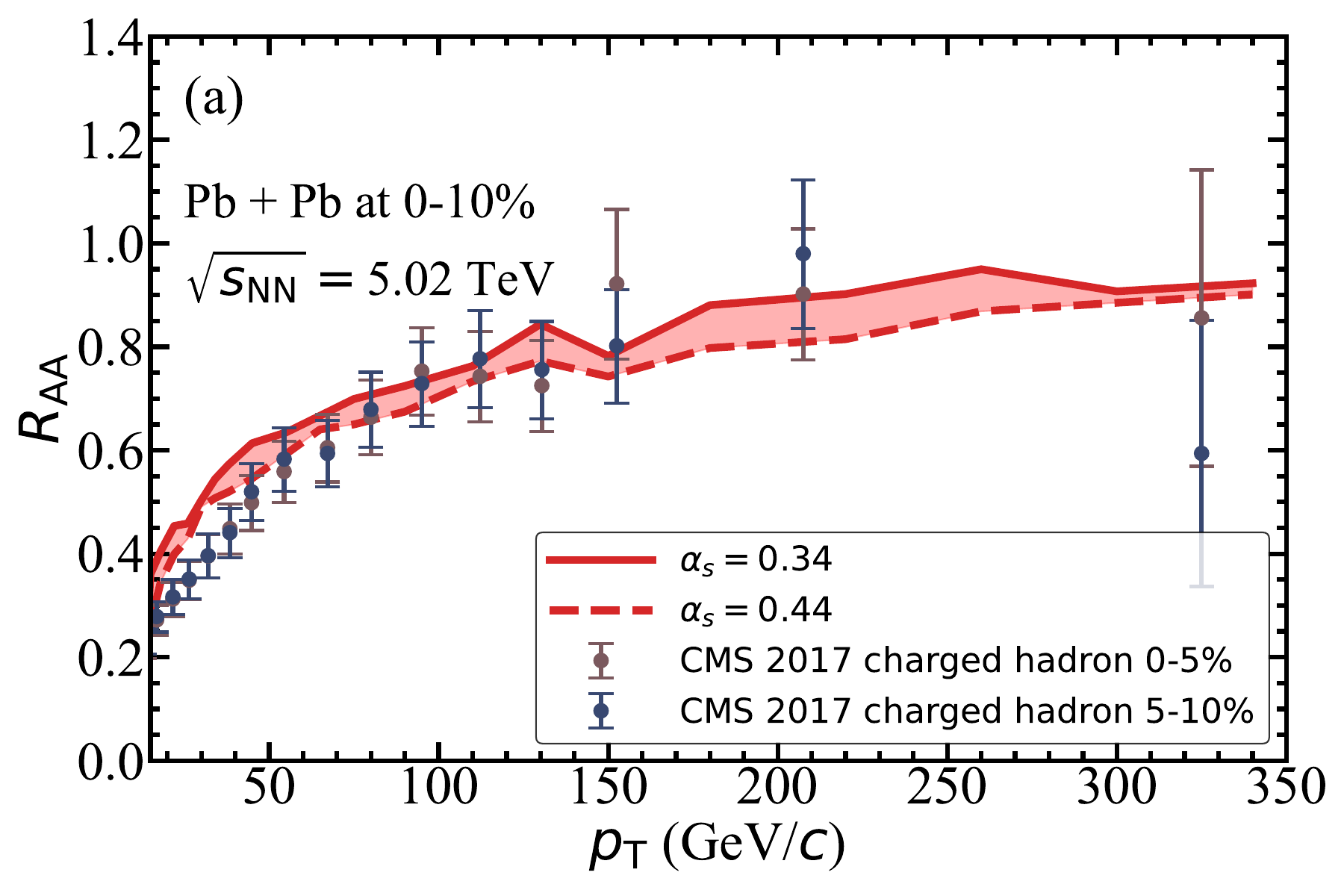}
    }
    \hfill 
    \hspace{-20pt}
    \subfigure{
        \label{fig:D0_hadron}
        \includegraphics[width=0.33\textwidth]{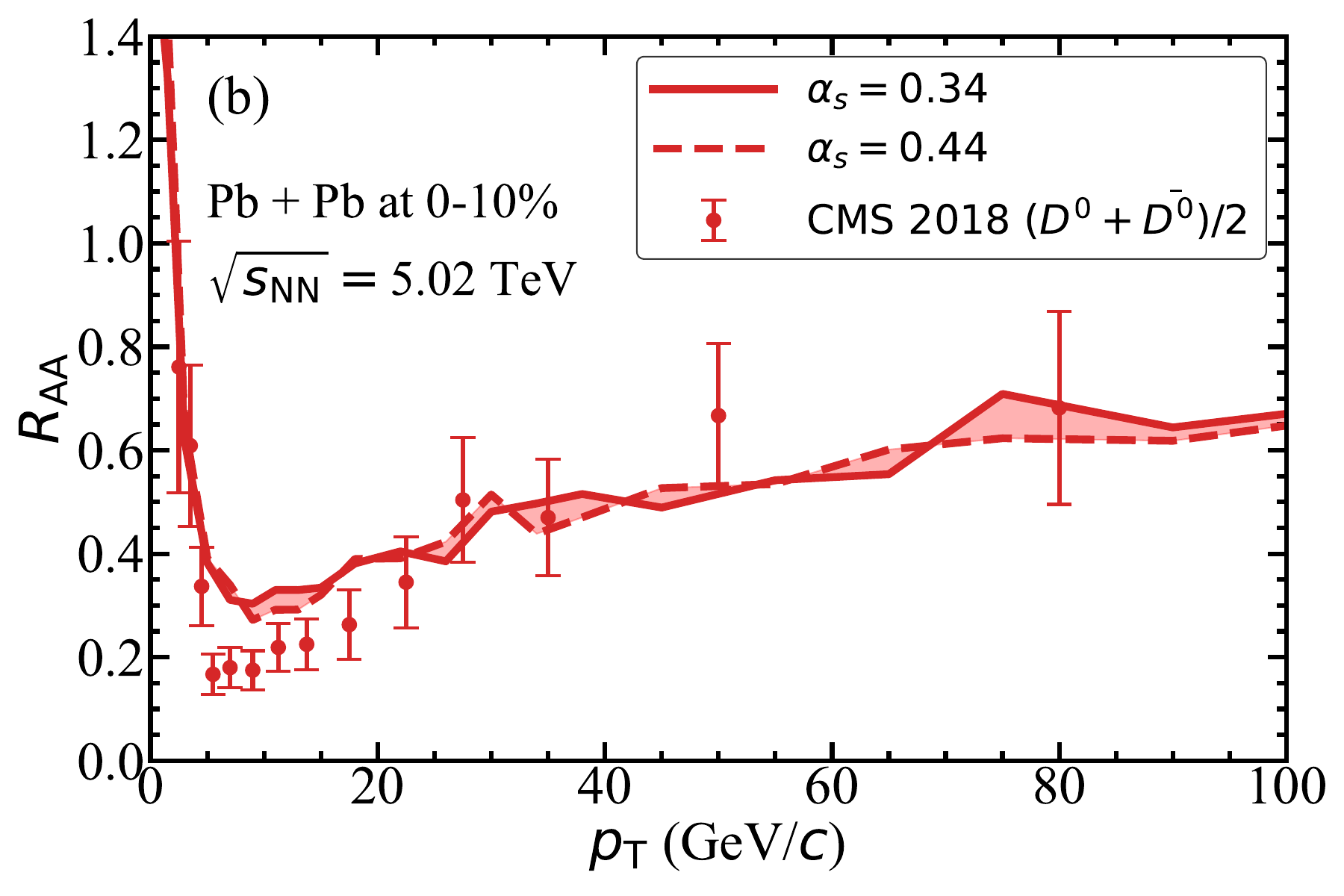}
    }
    \hfill
    \hspace{-20pt}
    \subfigure{
        \label{fig:Bpm_hadron}
        \includegraphics[width=0.33\textwidth]{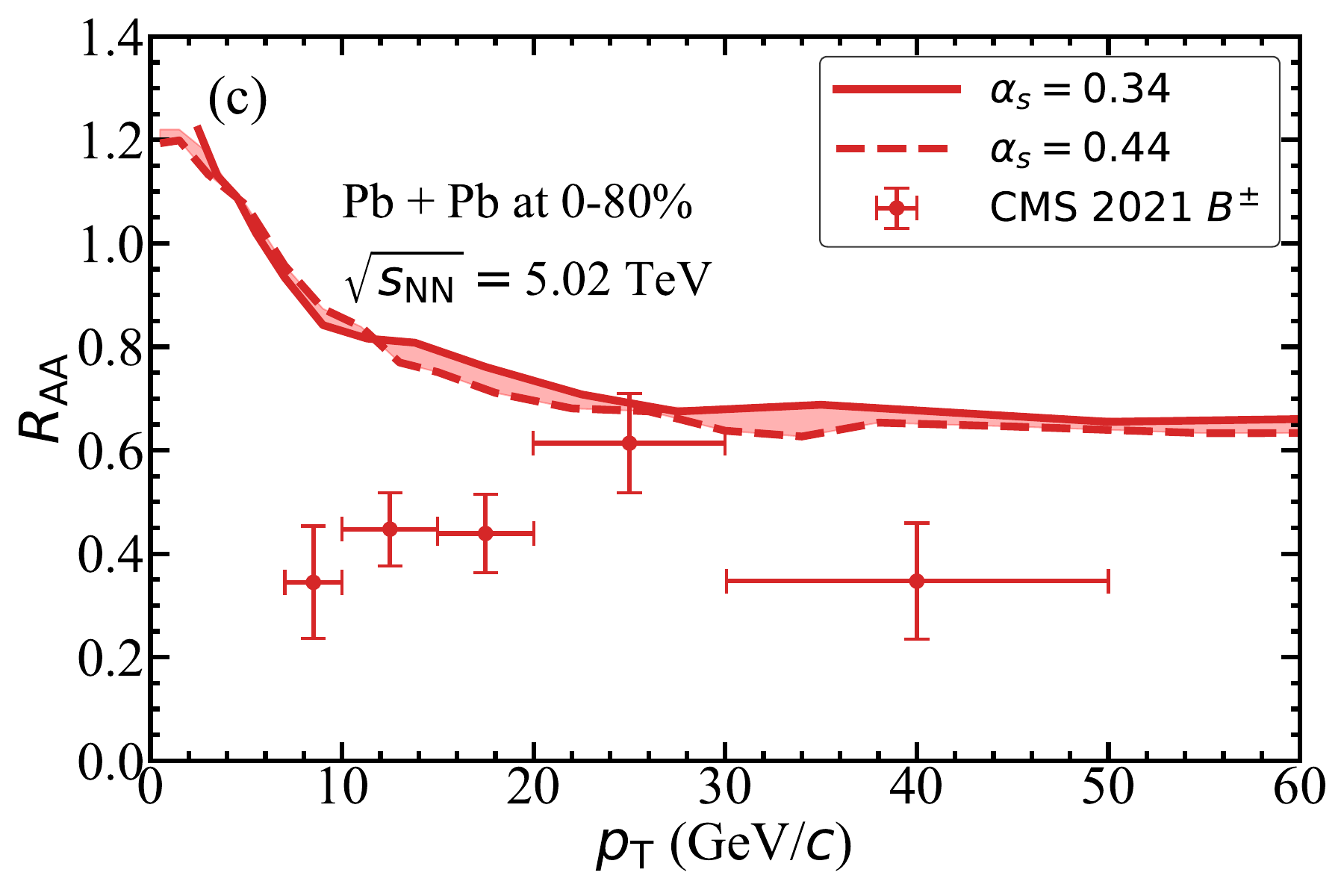}
    }
    
    \vspace{-10pt} % 行间距
    
    % 第二行：Jets
    \subfigure{
        \label{fig:charged_jet}
        \includegraphics[width=0.33\textwidth]{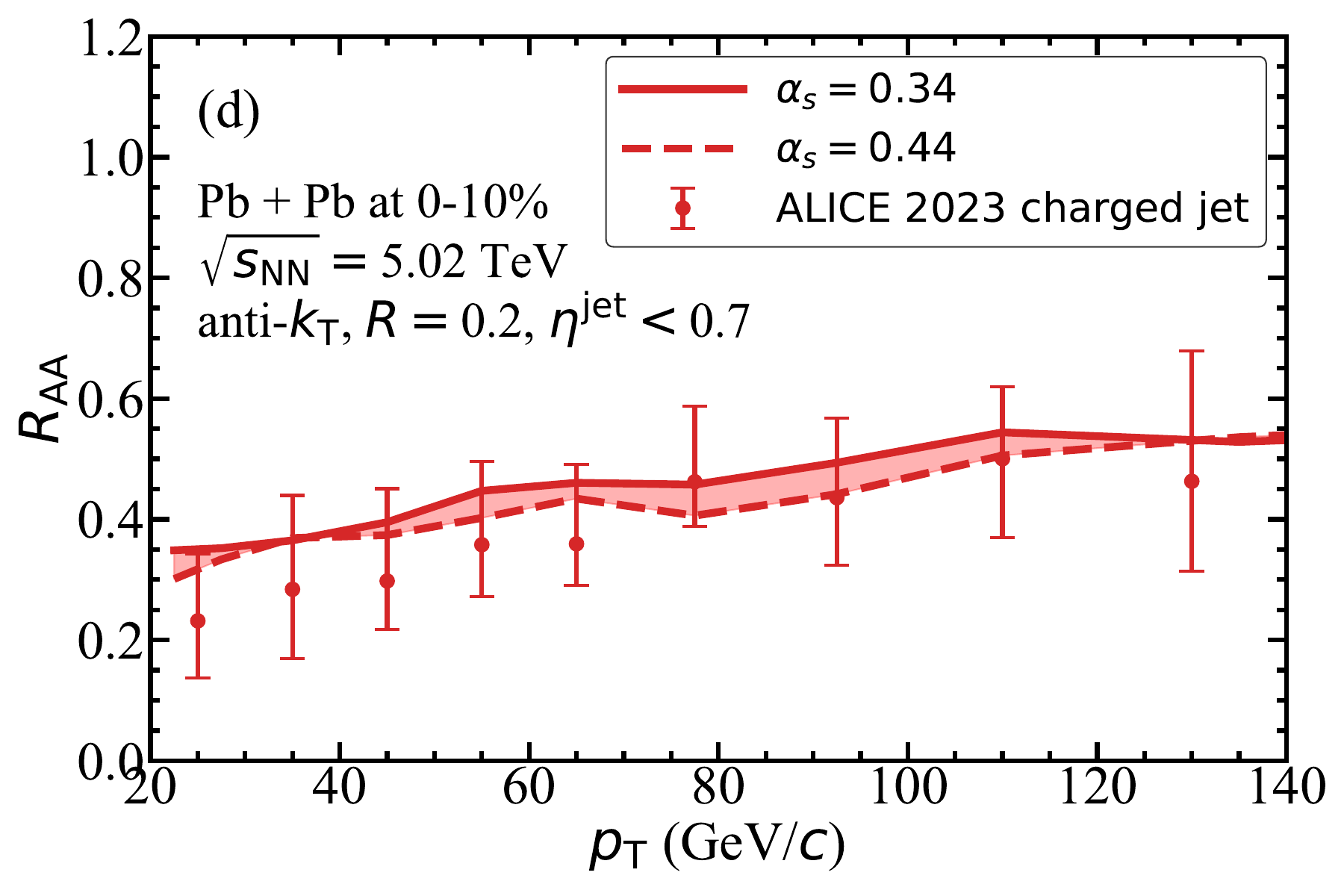}
    }
    \hfill
    \hspace{-20pt}
    \subfigure{
        \label{fig:c_jet}
        \includegraphics[width=0.33\textwidth]{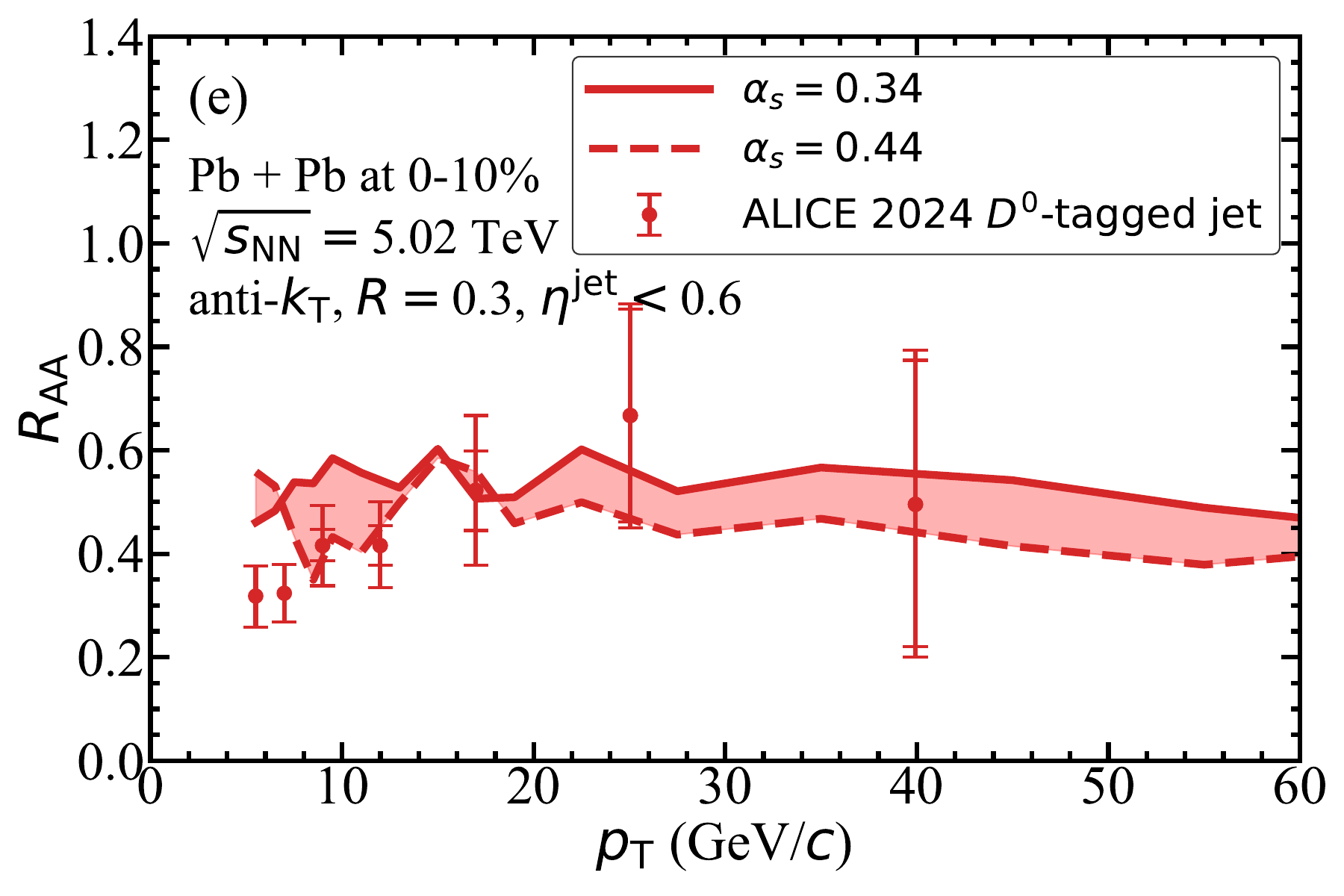}
    }
    \hfill
    \hspace{-20pt}
    \subfigure{
        \label{fig:b_jet}
        \includegraphics[width=0.33\textwidth]{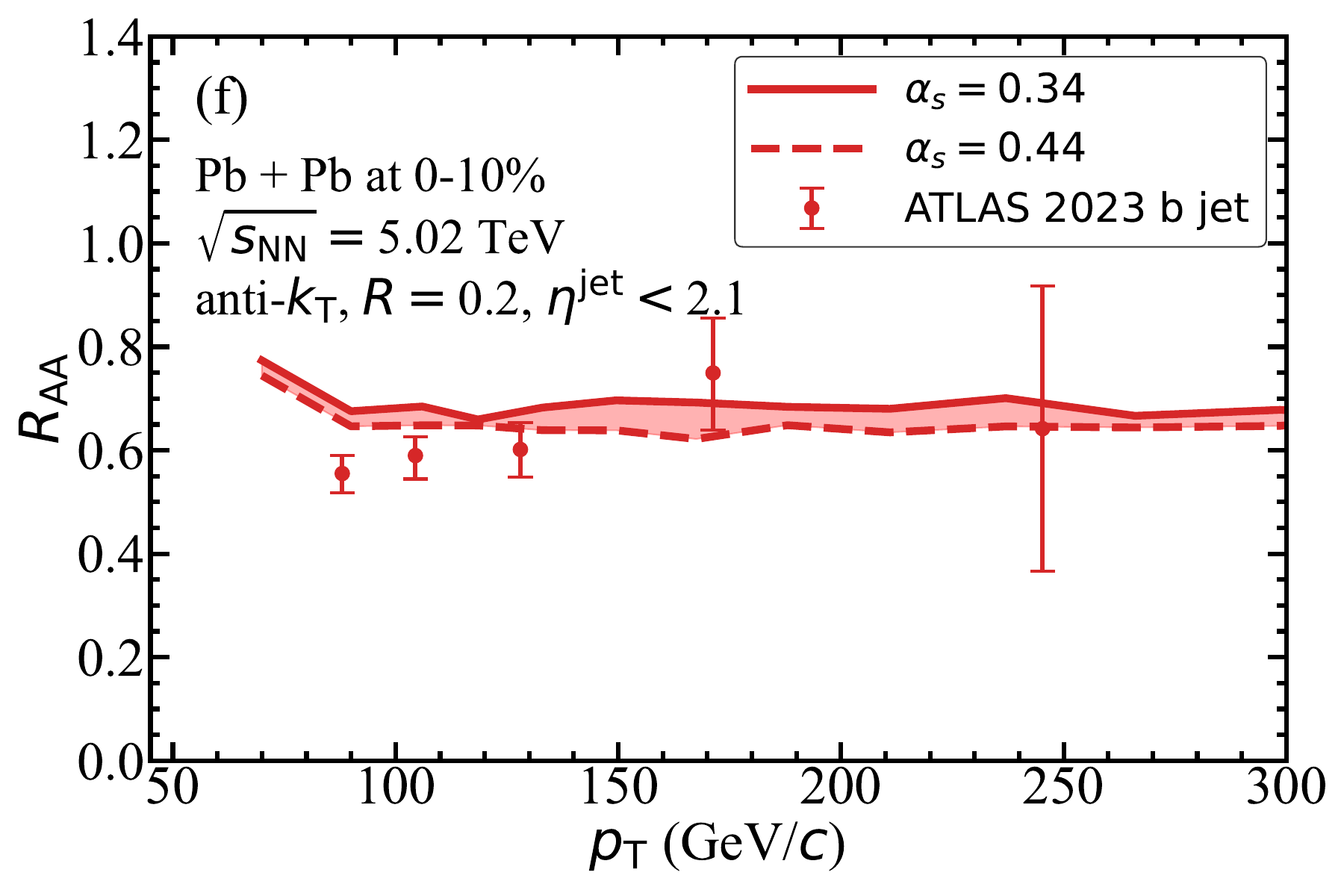}
    }
    
    \caption{(Color online) The nuclear modification factors of (a) charged hadrons, (b) $D$ mesons, (c) $B$ mesons, (d) charged jets, (e) $D^0$-tagged jets, and (f) $b$-tagged jets in Pb$+$Pb collisions at $\sqrt{s_{\mathrm{NN}}} = \SI{5.02}{TeV}$, compared to experimental data from the CMS~\cite{CMS:2021vui,CMS:2016xef,CMS:2017qjw,CMS:2017uoy}, ATLAS~\cite{Aaboud:2018twu,ATLAS:2022agz} and ALICE~\cite{ALICE:2023waz,Sheikh:2020sgf} Collaborations.}
    \label{fig:six_images}
\end{figure*}

There are no $2 \rightarrow 2$ scatterings in {Pythia} vacuum showers, and thus we design the color flow assignments for them as illustrated in Fig.~\ref{fig:scatter}. In each scattering digram, the upper left and right legs represent the incoming and outgoing jet partons, respectively. The lower left and right legs represent the initial (negative parton) and final (recoil parton) states of a medium constituent. The color flows are assigned such that the color charges are conserved between the initial and final states of each scattering, thereby allowing the jet partons produced by {Pythia~8} to remain connected by complete string configurations after evolving through the QGP. Note that the scattering matrices of these $2\rightarrow 2$ processes in Eq.~(\ref{eq:elRate}) are evaluated using the full color-flow information. The large-$N_c$ scheme is only applied when assigning numerical indices for the colors of partons, which is required for the vacuum showers and hadronization of these partons after they transition from LBT back to Pythia.

Because the contributions from negative partons must be subtracted from the positive parton contributions to jet observables, the two types need to be hadronized separately. Therefore, we replace the negative partons (lower left legs) in Fig.~\ref{fig:scatter} by ``fake" partons with negligible momenta ($p_x=p_y=p_z=0.1$~GeV) when conducting string fragmentation of positive partons. Finite, instead of zero, momenta are set for fake partons here for avoiding possible unphysical results from the {Pythia} string breaking. Similarly, finite masses, with default values in {Pythia~8}, are used for light flavor partons in hadronization. When being taken apart, negative partons cannot form color neutral strings by themselves, because they are produced independently at different locations inside the QGP. For this reason, we use the colorless hadronization scheme developed in Refs.~\cite{JETSCAPE:2019udz,Zhao:2020wcd,Zhao:2021vmu} to construct strings between negative partons based on minimizing their relative distances in the momentum space.

\section{Quenching of hadrons and jets}
\label{sec:mediumscale}

In this section, we first study how the medium scale $Q_\mathrm{M}$ affects the nuclear modification factors of hadrons and jets. Jets are reconstructed using hadrons via a modified {Fastjet} package~\cite{Cacciari:2011ma} in which the momenta of negative hadrons are subtracted from those of positive ones~\cite{He:2018xjv}. The nuclear modification factor is defined as the ratio of the hadron (or jet) spectrum in A$+$A collisions to that in $p+p$ collisions:
\begin{equation}
    R_{\mathrm{AA}}(p_{\mathrm{T}})\equiv\frac{\mathrm{d}N^{\mathrm{AA}}/\mathrm{d}p_{\mathrm{T}}}{\mathrm{d}N^{\mathrm{pp}}/\mathrm{d}p_{\mathrm{T}}\times\left\langle N_{\mathrm{coll}}^{\mathrm{AA}}\right\rangle},
\end{equation}
with $\left\langle N_{\mathrm{coll}}^{\mathrm{AA}}\right\rangle$ the average number of nucleon-nucleon binary collisions in each A$+$A collision.

Shown in Fig.~\ref{fig:QMeffect} are the nuclear modification factors of inclusive jets and charged hadrons in central (0-10\%) Pb$+$Pb collisions at $\sqrt{s_{\mathrm{NN}}}=\SI{5.02}{TeV}$, compared between using different values of the medium scale $Q_\mathrm{M}$. For each value of $Q_\mathrm{M}$, we adjust the strong coupling coefficient, the fixed $\alpha_\mathrm{s}$ parameter discussed in the previous section, to fit the jet $R_\mathrm{AA}$ data in Fig.~\ref{fig:QMeffect}(a), and examine how the hadron $R_\mathrm{AA}$ varies in Fig.~\ref{fig:QMeffect}(b). The choice of $Q_\mathrm{M} = \SI{0.5}{GeV}$ (red solid lines) corresponds to the settings in earlier versions of the LBT model, where jet partons evolve down to the hadronization scale $Q_\mathrm{h}$ in {Pythia} before entering LBT. This setup significantly underestimates the hadron $R_\mathrm{AA}$ data when $\alpha_\mathrm{s} = 0.1$ is tuned to describe the jet $R_\mathrm{AA}$ data. As $Q_\mathrm{M}$ is raised to 1~GeV (green dashed lines) and 2~GeV (blue dash-dotted lines), much better simultaneous descriptions of the jet and hadron $R_\mathrm{AA}$'s are achieved. In the LBT model, the energy loss of a high-energy parton depends weakly on its energy, leading to the following two effects when $Q_\mathrm{M}$ is raised. First, weak energy dependence of parton energy loss implies the energy loss of a full jet increases with the number of constituent partons inside the jet. A higher $Q_\mathrm{M}$ interrupts the initial stage of vacuum parton showers earlier, resulting in fewer jet constituent partons and thus reducing the jet energy loss. This explains why a larger $\alpha_\mathrm{s}$ is required to re-fit the jet $R_\mathrm{AA}$ in Fig.~\ref{fig:QMeffect}(a) when $Q_\mathrm{M}$ is raised. Second, a higher $Q_\mathrm{M}$ also causes less energy loss of leading partons. Vacuum parton showers can be effectively viewed as an ``energy loss" process for a leading parton, where the fractional energy loss -- represented by $z$ in the vacuum splitting function $P(z)$ -- depends weakly on the parton energy in the high-energy limit. Consequently, this ``energy loss" increases with the parton energy. Shifting the in-medium parton transport to a higher $Q_\mathrm{M}$ then reduces the parton energy before vacuum showers occur between $Q_\mathrm{M}$ and $Q_\mathrm{h}$, thereby decreasing the effective ``energy loss” in vacuum. On the other hand, this shift has only a mild effect on the in-medium energy loss of a leading parton, because the energy dependence of its in-medium energy loss is weak. These explain why the hadron $R_\mathrm{AA}$ in Fig.~\ref{fig:QMeffect}(b), which is driven by the leading parton energy loss, increases with $Q_\mathrm{M}$ even though a larger $\alpha_\mathrm{s}$ is applied at higher $Q_\mathrm{M}$. Although both jet and hadron suppression are reduced at higher $Q_\mathrm{M}$, the reduction is weaker for jets than for hadrons. One may understand this as follows. Raising $Q_\mathrm{M}$ extends the interaction time between jet partons and the QGP, and the extension is more significant for lower energy partons than for higher energy partons, because partons with higher energies usually form earlier and both the parton formation time ($\tau_\mathrm{f}$) and the QGP formation time ($\tau_0$) should be reached before a parton starts interacting with the QGP. Considering that the hadron suppression is dominated by the energy loss of leading partons, while the jet suppression is affected by medium modification of partons at different energies, increasing $Q_\mathrm{M}$ reduces suppression less for jets than for hadrons. Finally, as we re-fit the jet $R_\mathrm{AA}$ to data by increasing $\alpha_\mathrm{s}$, the hadron $R_\mathrm{AA}$ becomes larger. In other words, raising $Q_\mathrm{M}$ reduces the hadron-to-jet quenching ratio (or increases the hadron-to-jet $R_\mathrm{AA}$ ratio), improving the simultaneous description of the hadron and jet $R_\mathrm{AA}$'s.

We further present the nuclear modification factors of different species of hadrons and jets in Fig.~\ref{fig:six_images}. The left column show the $R_\mathrm{AA}$'s of (a) charged hadrons and (d) charged jets, the middle column for (b) $D^0$ ($\overline{D^0}$) mesons and (e) $D^0$-tagged jets, and the right column for (c) $B^{\pm}$ mesons and (f) $b$-tagged jets. We count jets that contain $B^{0}$, $B^{\pm}$ or $\Lambda_b$ hadrons as $b$-tagged jets in our calculation. Here, we use the medium scale at $Q_\mathrm{M} = \SI{2}{GeV}$ and show the uncertainty bands for $\alpha_\mathrm{s}$ between 0.34 and 0.44, illustrating the sensitivity of the hadron and jet $R_\mathrm{AA}$'s to the strong coupling coefficient. In general, our improved LBT model provides a consistent description of the nuclear modification factors across hadrons and jets with different flavors. The remaining deviations from the experimental data, especially for $B$ mesons, mainly result from the rough treatment of the non-perturbative effects at low momenta. Rigorously, non-perturbative effects should be stronger for partons with heavier masses~\cite{Dang:2023tmb}, which may not be approximated by the same $K_p$ factor across all flavors of partons. In addition, while the string fragmentation effectively contains the coalescence between jet partons (including emitted gluons and recoil partons) in hadronization, coalescence between jet partons and the QGP constituents has not been fully incorporated, which should have a sizable impact on the hadron spectra at low to intermediate $p_\mathrm{T}$~\cite{Zhao:2021vmu,Cao:2019iqs,Zhao:2023nrz}. For these reasons, we confine our study to understanding the uncertainties introduced by different model parameters, rather than to precisely extracting their values.

\begin{figure}[tbp!]
    \centering
    \includegraphics[width=0.95\linewidth]{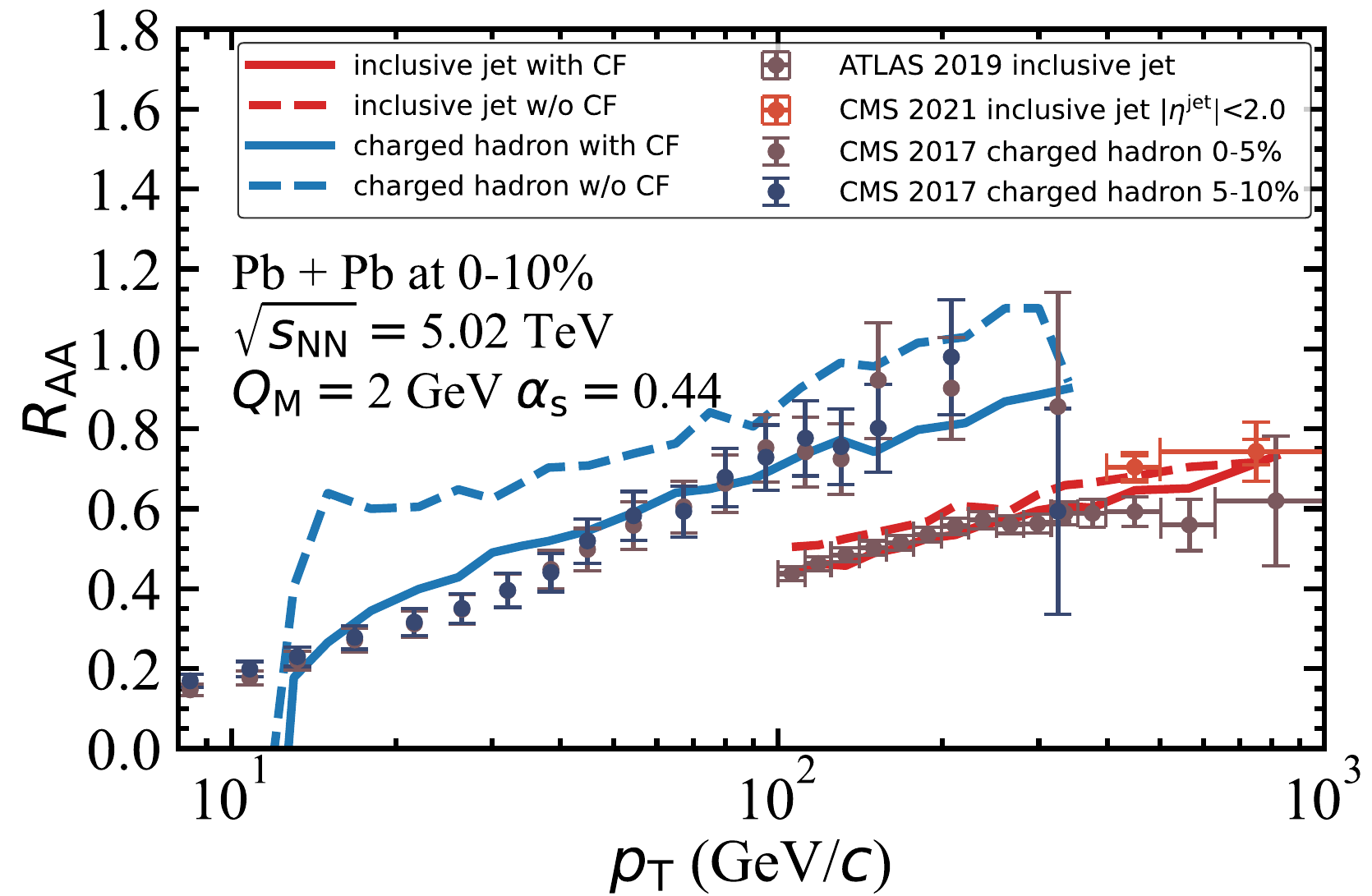} 
    \caption{(Color online) The nuclear modification factors of inclusive jets and hadrons in central Pb$+$Pb collisions at $\sqrt{s_{\mathrm{NN}}}=\SI{5.02}{TeV}$, compared between including and not including the color flow (CF) information in the Pythia string fragmentation. The experimental data are taken from the CMS~\cite{CMS:2021vui,CMS:2016xef} and ATLAS~\cite{Aaboud:2018twu} Collaborations.}
    \label{fig:colorless}
\end{figure}

In the end, we investigate how the string fragmentation is affected by the color flow configurations of jet partons. In Fig.~\ref{fig:colorless}, we compare the $R_{\mathrm{AA}}$'s of hadrons and jets between including and not including color flow (CF) information in the Pythia string fragmentation. The former, shown by the solid lines, corresponds to our default setting, in which the color configurations of positive partons are correlated with the medium-modified parton shower history, as described in Sec.~\ref{sec:model}. For the latter, shown by the dashed lines, we use the colorless hadronization model~\cite{Zhao:2021vmu} to convert positive partons into hadrons, in which the colors of positive partons are re-assigned based on their relative distances in momentum space upon hadronization. Negative partons are hadronized using the colorless model in both cases. While these two setups provide similar jet $R_{\mathrm{AA}}$'s, the hadron $R_{\mathrm{AA}}$ obtained from the colorless hadronization model is apparently larger than that from the color-tracking model. Compared to our color-tracking model where a leading parton can be frequently connected to a medium response particle at the thermal energy scale, a leading parton in the colorless hadronization model is connected to a nearby parton in the momentum space. It is likely that the latter configuration produces hadrons with higher energies via string breaking. Hence, for a fixed parton sample, color configurations affect the momentum distribution of hadrons. Although the total jet energy remains largely unaffected, we expect that the jet fragmentation function should also depend on this color configuration. There are other color flow channels for $2\rightarrow 2$ scatterings besides those presented in Fig.~\ref{fig:scatter}. We have verified that varying the assumed color configurations for $t$-channel scatterings -- for instance, by connecting the red lines between the top-left and bottom-left partons and the green lines between the top-right and bottom-right partons in the upper row of Fig.~\ref{fig:scatter} -- has negligible impact on the $R_\mathrm{AA}$'s of hadrons and jets. A more rigorous treatment, however, would necessitate implementing all possible color flow channels weighted by their individual probabilities, a refinement planned for future work.

\section{Summary}
\label{sec:summary}

We improve the LBT model to enable a simultaneous description of the nuclear modification factors of hadrons and jets. A medium virtuality scale ($Q_{\mathrm{M}}$) is introduced, at which the Pythia vacuum showers are interrupted by the LBT model for parton-QGP interactions. The vacuum showers resume when jet partons exit the QGP, and evolve jet partons to the hadronization scale. We introduce color flows for both elastic and inelastic scatterings, allowing jet partons, medium-induced gluons, and medium response particles to be connected by strings that encode the information of the medium-modified parton shower history. Final state partons are converted into hadrons via the Pythia string fragmentation.

Within this improved framework, we find that raising $Q_{\mathrm{M}}$ reduces the quenching of both hadrons and jets if $\alpha_\mathrm{s}$ is fixed. However, this reduction is weaker for jets than for hadrons. Raising $Q_{\mathrm{M}}$ extends the interaction time between jet partons and the QGP. Since softer jet partons typically form later than harder ones, this extension enhances the medium modification of softer partons more than that of harder ones. Because the high-$p_\mathrm{T}$ hadron spectrum is dominated by leading partons in jets, while the full jet spectrum depends on both hard and soft components, a larger $Q_{\mathrm{M}}$ leads to a larger ratio of jet to hadron quenching, or a smaller $R_\mathrm{AA}$ ratio between jets and hadrons. By adjusting $\alpha_\mathrm{s}$, a simultaneous description of hadron and jet quenching can be achieved for $Q_{\mathrm{M}}$ around 1 to $\SI{2}{GeV}$. Furthermore, in string fragmentation, hadrons are found to be more sensitive to partonic color configurations than jets are. While including and not including the parton-shower-correlated color flow information in the Pythia string fragmentation yield comparable jet $R_\mathrm{AA}$'s, the hadron $R_\mathrm{AA}$ obtained when the color-flow information is included is clearly smaller than that obtained when it is not included.

Overall, the improved LBT model provides a consistent description of the suppression factors of hadrons and jets across different flavor tags and also establishes a baseline for quantitatively understanding of the nuclear modification of hadron-triggered jets, as reported in a separate study~\cite{Jing:2025bwi}. The source code of this model is publicly available in Ref.~\cite{githubLBT}. Further developments could incorporate more realistic treatments of non-perturbative corrections to parton scatterings at low momenta, especially for heavy quarks, and take into account coalescence between jet partons and the QGP constituents in hadronization. Bayesian inference should also be implemented for a more precise calibration of the model parameters, in particular $Q_\mathrm{M}$, an intrinsic property of the QGP.

\section*{Acknowledgments}
We are grateful for valuable discussions with Lejing Zhang, Changle Sun, and Peng Jing. This work is supported by the National Natural Science Foundation of China (NSFC) under Grant Nos.~12575146, 12175122, 2021-867, 12505154, 12225503, and~12321005.

\end{document}